\documentclass[sigconf]{acmart}
\usepackage{tikz}
\usetikzlibrary{shapes.geometric, arrows, positioning, fit, backgrounds}
\usepackage{graphicx}
\usepackage{subcaption}

\AtBeginDocument{%
  }

\copyrightyear{2026}
\acmYear{2026}
\setcopyright{cc}
\setcctype{by}
\acmConference[CHI EA '26]{Extended Abstracts of the 2026 CHI Conference on Human Factors in Computing Systems}{April 13--17, 2026}{Barcelona, Spain}
\acmBooktitle{Extended Abstracts of the 2026 CHI Conference on Human Factors in Computing Systems (CHI EA '26), April 13--17, 2026, Barcelona, Spain}
\acmDOI{10.1145/3772363.3798823}
\acmISBN{979-8-4007-2281-3/2026/04}

\begin{document}

\title{Grant, Verify, Revoke: A User-Centric Pattern for Blockchain Compliance}

\author{Supriya Khadka}
\affiliation{%
  \institution{George Mason University}
  \city{Fairfax, Virginia}
  \country{USA}}
\email{skhadk@gmu.edu}

\author{Sanchari Das}
\affiliation{%
  \institution{George Mason University}
  \city{Fairfax, Virginia}
  \country{USA}}
\email{sdas35@gmu.edu}

\renewcommand{\shortauthors}{Khadka and Das}

\begin{abstract}
In decentralized web applications, users face an inherent conflict between public verifiability and personal privacy. To participate in regulated on-chain services, users must currently disclose sensitive identity documents to centralized intermediaries, permanently linking real-world identities to public transaction histories. This binary choice between total privacy loss or total exclusion strips users of agency and exposes them to persistent surveillance. In this work, we introduce a~\emph{Selective Disclosure Framework} designed to restore user sovereignty by decoupling eligibility verification from identity revelation. We present~\emph{ZK-Compliance}, a prototype that leverages browser-based zero-knowledge proofs to shift the interaction model, enabling users to prove specific attributes (e.g., ``I am over 18'') locally without revealing the underlying data. We implement a user-governed~\emph{Grant, Verify, Revoke} lifecycle that transforms the user's mental model of compliance from a permanent data handover into a dynamic, revocable authorization session. Our evaluation shows that client-side proof generation takes under $200~ms$, enabling a seamless interactive experience on commodity hardware. This work provides early evidence that regulatory compliance need not come at the cost of user privacy or autonomy.
\end{abstract}



\maketitle

\section{Introduction}
A fundamental tension exists between the Web3 promise of self-sovereignty and the operational reality of radical transparency. While public blockchains enable trustless verification, they force users into a hostile environment where every interaction is permanently recorded in a globally visible ledger~\cite{conti2018survey, grover2025sok, kishnani2023blockchain, agarwal2025systematic,huang2025systemization,borisov2021financial,werbach2018trust}. As these platforms expand into regulated domains like finance and gaming, users face a structural conflict between participation and privacy~\cite{zachariadis2019governance, rishiwal2024blockchain}. Currently, to satisfy identity-based controls such as Know Your Customer (KYC) requirements~\cite{fatf2019vasp}, users are forced to rely on binary disclosure models that fail to capture the nuance of social trust~\cite{khadka2026poster, tan2023trust}.

The core user experience challenge stems from the immutability of on-chain data. Because public blockchains cannot be redacted, users risk falling into a~\textit{transparency trap} where complying with a simple verification request (e.g., proving age) permanently binds their real-world identity to their entire transaction history~\cite{meiklejohn2013fistful, buterin2016privacy, agarwal2025systematic}. This linkage enables surveillance and profiling that directly conflict with data minimization principles~\cite{gdpr2016}. Recent user-centered research argues that privacy must be instantiated as a controllable interaction mechanism rather than a static back-end policy~\cite{sharma2024m,adhikari2025natural,adhikari2025policypulse}. Our framework adopts this stance, elevating selective disclosure from a cryptographic possibility to a first-class user capability.

Currently, most applications address this tension by outsourcing verification to centralized intermediaries~\cite{belchior2021survey, klems2017trustless}. While this approach satisfies regulatory requirements, it reintroduces the very centralization risks that blockchain systems aim to avoid like data breaches and censorship~\cite{kishnani2022privacy, kishnani2023assessing}. Empirical studies confirm that this model generates significant \textit{privacy discomfort}, with users expressing anxiety over unauthorized data retention and the lack of recourse against service denial~\cite{si2024understanding, conti2018survey}. This centralization effectively strips users of agency, reducing them to passive subjects of verification rather than active participants.

To address this, in this paper, we propose a \emph{Selective Disclosure Framework} that resolves the structural tension between compliance and user agency. We argue that privacy should be an active capability of the user. The framework leverages Zero-Knowledge Proofs (ZKPs), specifically zk-SNARKs~\cite{ben2014succinct}, to allow users to prove compliance predicates on-chain without revealing underlying identity documents. Rather than treating identity as a monolithic object to be surrendered, our framework enforces fine-grained, user-controlled disclosure. We instantiate this through~\emph{ZK-Compliance}, a prototype that enables privacy-preserving age verification via a client-side model. The system introduces an explicit~\emph{Grant, Verify, Revoke} lifecycle~\cite{khadka2026poster}, reframing access authorization as a temporary, revocable state rather than a permanent data transfer. This design restores user control while preserving the composability required by blockchain applications.

\section{The Agency Gap in Web3 Identity}
Research on blockchain identity often bifurcates into transactional privacy protocols, decentralized identity frameworks, and hardware-based controls~\cite{Bernabe2019, garcia2024surveyblockchainbasedprivacyapplications}. We analyze this landscape through the lens of Human-Computer Interaction (HCI), focusing on the cognitive load of compliance and the limitations of current trust models.

Early privacy tools, such as Zerocash~\cite{sasson2014zerocash}, focused on obfuscating financial metadata by hiding senders, recipients, and values. While effective for financial confidentiality, these ``all-or-nothing'' models fail to address identity-based access control. From a user perspective, these tools lack granularity: a user cannot prove they are ``over 18'' without either revealing their entire financial history or nothing at all. This leaves users without the means to satisfy narrow eligibility predicates, forcing a choice between total transparency (compliance) and total opacity (non-compliance). Users require attribute-based mechanisms that separate eligibility verification from financial disclosure.

Attempts to solve this through the Self-Sovereign Identity (SSI) paradigm and W3C Verifiable Credentials~\cite{w3c2022vc} have introduced significant interaction friction. SSI implementations often require users to navigate complex trust stacks involving Decentralized Identifiers (DIDs) and specialized wallet software, creating a high barrier to entry~\cite{Krul_2024, Ishmaev2021, yu2024eggsbasket}. This disconnect between the theoretical promise of self-sovereignty and the practical difficulty of key management leads to ``security fatigue,'' often causing users to revert to custodial solutions for convenience~\cite{sharma2024can}. Critically, most SSI interactions act as ``stateless events'', where once a user presents a credential, they typically lose control over how that data is stored or used~\cite{bruhner2023bridging}. This results in a one-way disclosure rather than a managed relationship.

Finally, approaches using Trusted Execution Environments (TEEs) like Enigma~\cite{zyskind2015enigmadecentralizedcomputationplatform} and Ekiden~\cite{Cheng_2019} attempt to bridge this gap by offloading computation to secure hardware. However, these systems introduce a ``black box'' trust model, asking users to rely on opaque hardware guarantees (e.g., Intel SGX) that they cannot audit or verify. This opacity complicates the user's mental model of trust and introduces vulnerability to side-channel attacks~\cite{zhang2019security, das2020mfa}. In contrast, our framework minimizes reliance on external hardware trust, using lightweight cryptographic circuits that execute transparently within the user's standard web browser.

\section{System Design: A User-Centric Lifecycle}
\begin{figure}[ht]
    \centering
    \definecolor{grantGreen}{RGB}{20, 110, 40}   
    \definecolor{activeOrange}{RGB}{230, 126, 34} 
    \definecolor{revokeRed}{RGB}{192, 57, 43}     
    \definecolor{chainBlue}{RGB}{52, 152, 219}    

    \resizebox{\linewidth}{!}{%
    \begin{tikzpicture}[
        node distance=1.5cm and 1.2cm,
        auto,
        every node/.style={font=\sffamily\small},
        usernode/.style={rectangle, minimum width=2.5cm, minimum height=1.2cm, align=center, draw=grantGreen, very thick, fill=white, rounded corners=4pt},
        process/.style={rectangle, minimum width=2.2cm, minimum height=1cm, align=center, draw=gray!70, fill=white, rounded corners, thick},
        contract/.style={rectangle, minimum width=2.2cm, minimum height=1cm, align=center, draw=chainBlue!80!black, fill=chainBlue!10, thick},
        appnode/.style={rectangle, minimum width=2.2cm, minimum height=1cm, align=center, draw=activeOrange, fill=activeOrange!10, thick, rounded corners},
        arrow/.style={thick,->,>=stealth, rounded corners, font=\sffamily\footnotesize\bfseries},
        labelbox/.style={font=\bfseries\footnotesize, text=gray!70!black, align=left}
    ]

    \node (user) [usernode] {Alice (User) \\ Secure Vault};
    \node (prover) [process, right=1.2cm of user] {Local ZK Prover\\(Browser Module)};
    \node (proof) [process, right=1.8cm of prover] {Ephemeral Proof\\(Anonymous Token)};

    \node (verifier) [contract, below=1.8cm of proof] {Verifier\\Contract};
    \node (registry) [contract, left=1.8cm of verifier] {Access\\Registry};

    \node (dapp) [appnode, left=2.2cm of registry] {TradeBase App};


    \draw [arrow, grantGreen] (user) -- node[above, align=center, scale=0.8] {1. Grants\\Consent} (prover);
    \draw [arrow, grantGreen] (prover) -- node[above, align=center, scale=0.8] {2. Generates Proof\\(Locally)} (proof);
    \draw [arrow, grantGreen] (proof) -- node[right, scale=0.8, align=left] {3. Submits\\to Chain} (verifier);
    \draw [arrow, grantGreen] (verifier) -- node[above, scale=0.8, align=center] {4. Activates \\ Session} (registry);

    \draw [arrow, activeOrange, dashed] (dapp.east) to[bend right=45] node[right, scale=0.8, pos=0.3, yshift=-0.3cm, xshift=-0.6cm] {5. Verify Status} (registry.west);
    \draw [arrow, activeOrange, dashed] (registry.west) to[bend right=45] node[left, scale=0.8, pos=0.5, yshift=0.25cm, xshift=0.9cm] {6. Access Granted} (dapp.east);

    \draw [arrow, revokeRed, very thick] (user) -- node[left, scale=0.8, pos=0.45, xshift=-10pt, align=right] {7. Alice Triggers\\\textbf{'Kill Switch'} (Revoke)} (registry);

    \begin{pgfonlayer}{background}
        \node [fit=(user) (prover) (proof), draw=grantGreen!30, fill=grantGreen!5, rounded corners, dashed, inner sep=0.4cm] (clientbox) {};
        \node [above right, xshift=-5pt] at (clientbox.north west) [labelbox, text=grantGreen!60!black] {Alice's Sovereign Domain (Client Side)};

        \node [fit=(verifier) (registry), draw=chainBlue!30, fill=chainBlue!5, rounded corners, inner sep=0.4cm] (chainbox) {};
        \node [above right, xshift=5pt] at (chainbox.north west) [labelbox, text=chainBlue!70!black] {Ethereum (Public State)};
    \end{pgfonlayer}

    \end{tikzpicture}
    }
    \caption{The \emph{Grant, Verify, Revoke} lifecycle. (1-4) \textbf{Grant:} Alice actively generates a private proof locally to establish a session. (5-6) \textbf{Verify:} The application continuously checks this active, time-bounded authorization. (7) \textbf{Revoke:} Alice retains a direct "Kill Switch," allowing her to instantly sever the connection on-chain, bypassing the application entirely.}
    \label{fig:architecture_lifecycle}
\Description[A user-centric architecture diagram showing the interaction flow between Alice, the Ethereum Blockchain, and the TradeBase app.]{The diagram illustrates the 'Grant, Verify, Revoke' lifecycle, emphasizing user agency.
1. Grant Phase (Green arrows): Alice, from her 'User & Secure Vault', takes action '1. Alice Grants Consent' leading to local proof generation and submission to the chain, ending with '4. Activates Session' in the Access Registry.
2. Verify Phase (Orange dashed arrows): The 'TradeBase App' continuously interacts with the registry to '5. Verify Status' and confirm '6. Access Granted'.
3. Revoke Phase (Bold Red arrow): A direct, thick red line labeled '7. Alice Triggers 'Kill Switch' (Revoke)' connects Alice directly to the Access Registry, showing her ability to override the entire system instantly.}
\end{figure}
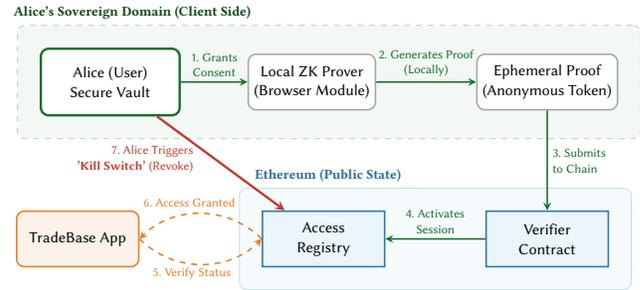

We propose a system architecture designed to align with the user's mental model of~\textit{temporary permission} rather than~\textit{permanent submission}. Traditional compliance workflows rely on a custodial mental model: users hand over raw documents (like a passport) to a centralized server, losing all control over how that data is stored or when it is deleted~\cite{sharma2024can}. Our architecture shifts this to a sovereign mental model. It coordinates three actors: \textbf{User} (Prover), \textbf{Issuer} (Trust Root), and \textbf{Verifier} (Service Provider). The system enforces a strict separation between \textit{private attribute storage} and \textit{public authorization state}. To operationalize this, we introduce a \emph{Grant, Verify, Revoke} lifecycle. This model transforms the compliance process from a static data hand-off into a dynamic, revocable session. The architecture is visualized in Figure~\ref{fig:architecture_lifecycle}.

\paragraph{\textbf{Phase 1: Grant (Sovereign Proving)}}
The lifecycle begins with the user's~\textit{Identity Vault}, a secure, client-side storage layer residing entirely within the user's browser or device. Consider a standard usage scenario where~\textit{Alice} wants to access an age-restricted trading platform. In a traditional system,~\textit{Alice} would upload a photo of her ID card. In our system, the platform requests an eligibility check. The local module retrieves her birthdate from her vault and executes the verification circuit locally. This phase is key for user agency as it shifts the~\textit{moment of verification} from a third-party server to the user's own device. By generating the zero-knowledge proof locally,~\textit{Alice} acts as the sole custodian of her data. The output is a cryptographic proof that asserts the truth of the predicate (e.g., ``Age > 18'') without revealing her actual birthdate. The network learns~\textit{about} the user without learning the user's data.

\begin{figure*}[htbp]
    \centering
    \begin{subfigure}[b]{0.48\textwidth}
        \centering
        \includegraphics[width=\linewidth]{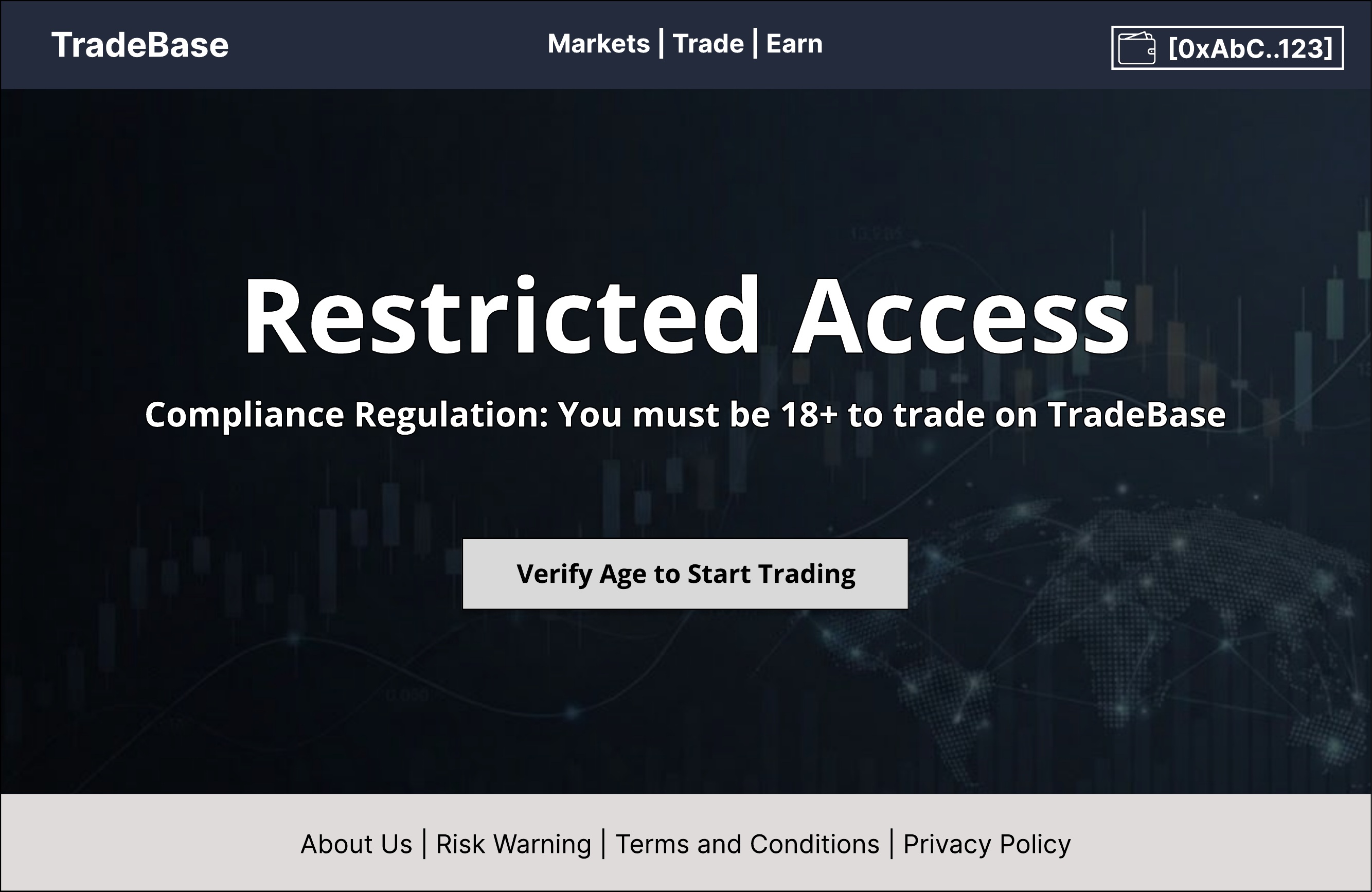}
        \caption{Restricted Access (TradeBase Host)}
        \label{fig:page1}
    \end{subfigure}
    \hfill
    \begin{subfigure}[b]{0.48\textwidth}
        \centering
        \includegraphics[width=\linewidth]{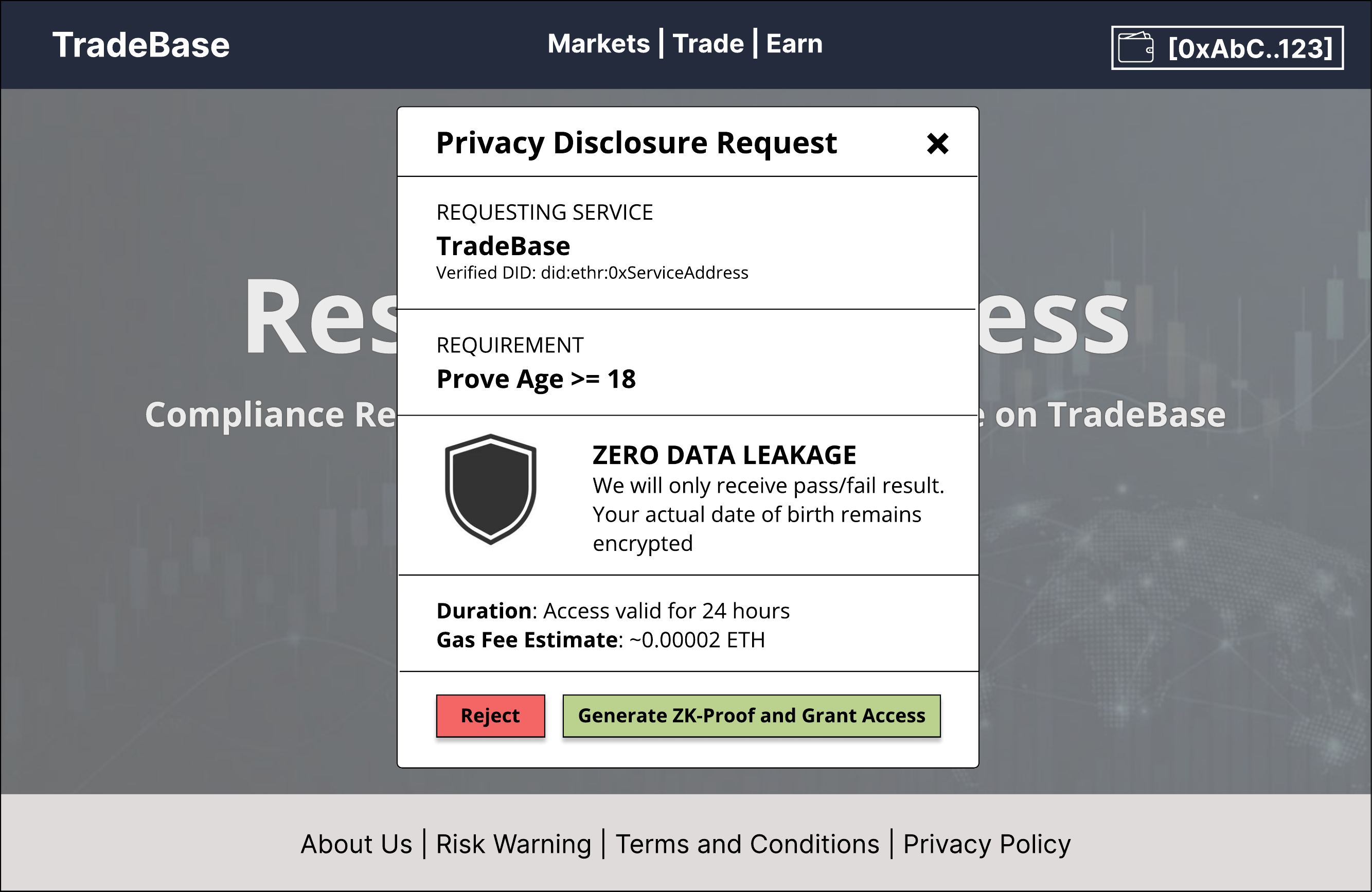}
        \caption{ZK-Compliance Consent Request}
        \label{fig:page2}
    \end{subfigure}
    
    \vspace{0.5cm}

    \begin{subfigure}[b]{0.48\textwidth}
        \centering
        \includegraphics[width=\linewidth]{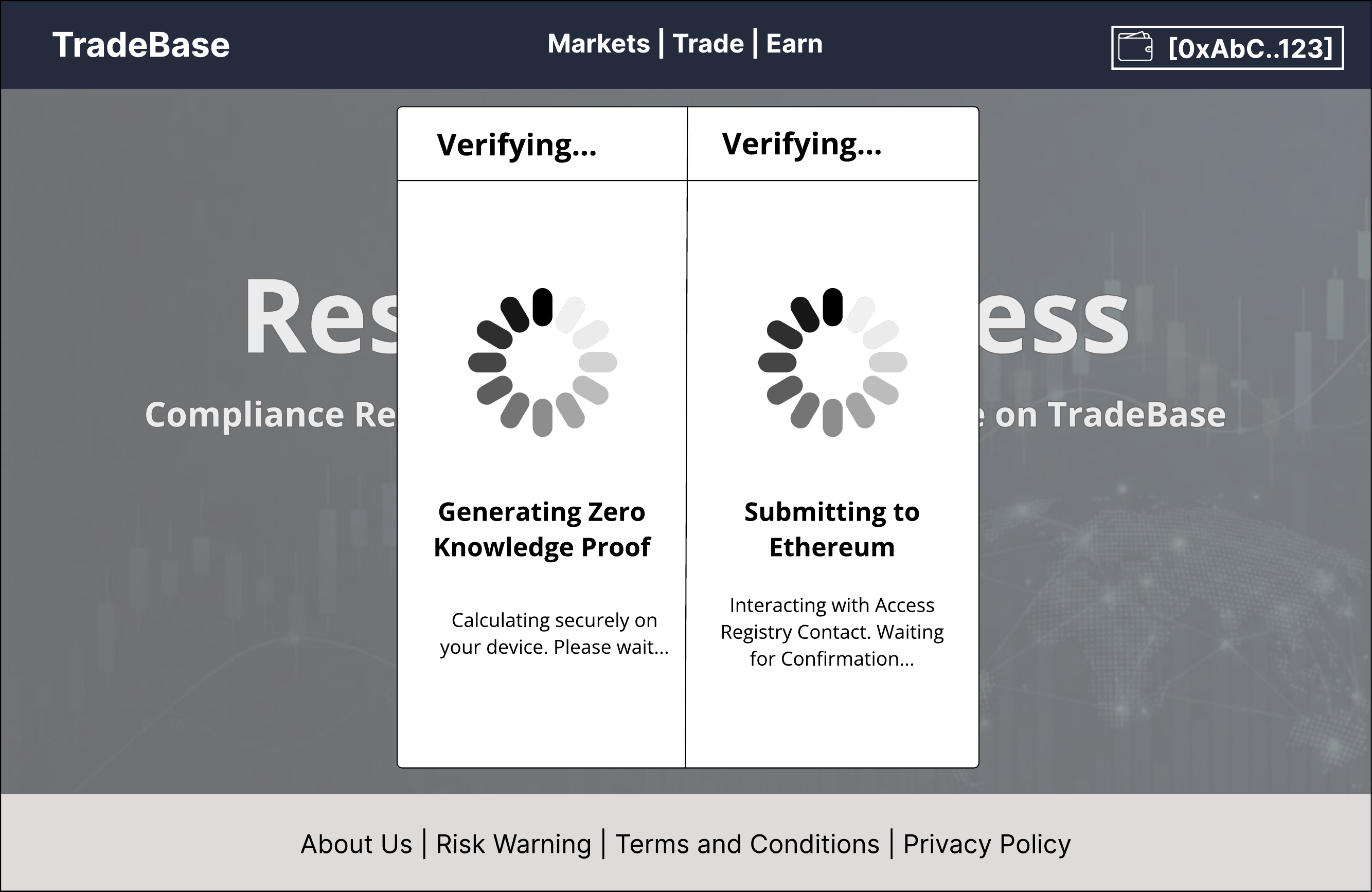}
        \caption{Client-Side Proof Generation}
        \label{fig:page3}
    \end{subfigure}
    \hfill
    \begin{subfigure}[b]{0.48\textwidth}
        \centering
        \includegraphics[width=\linewidth]{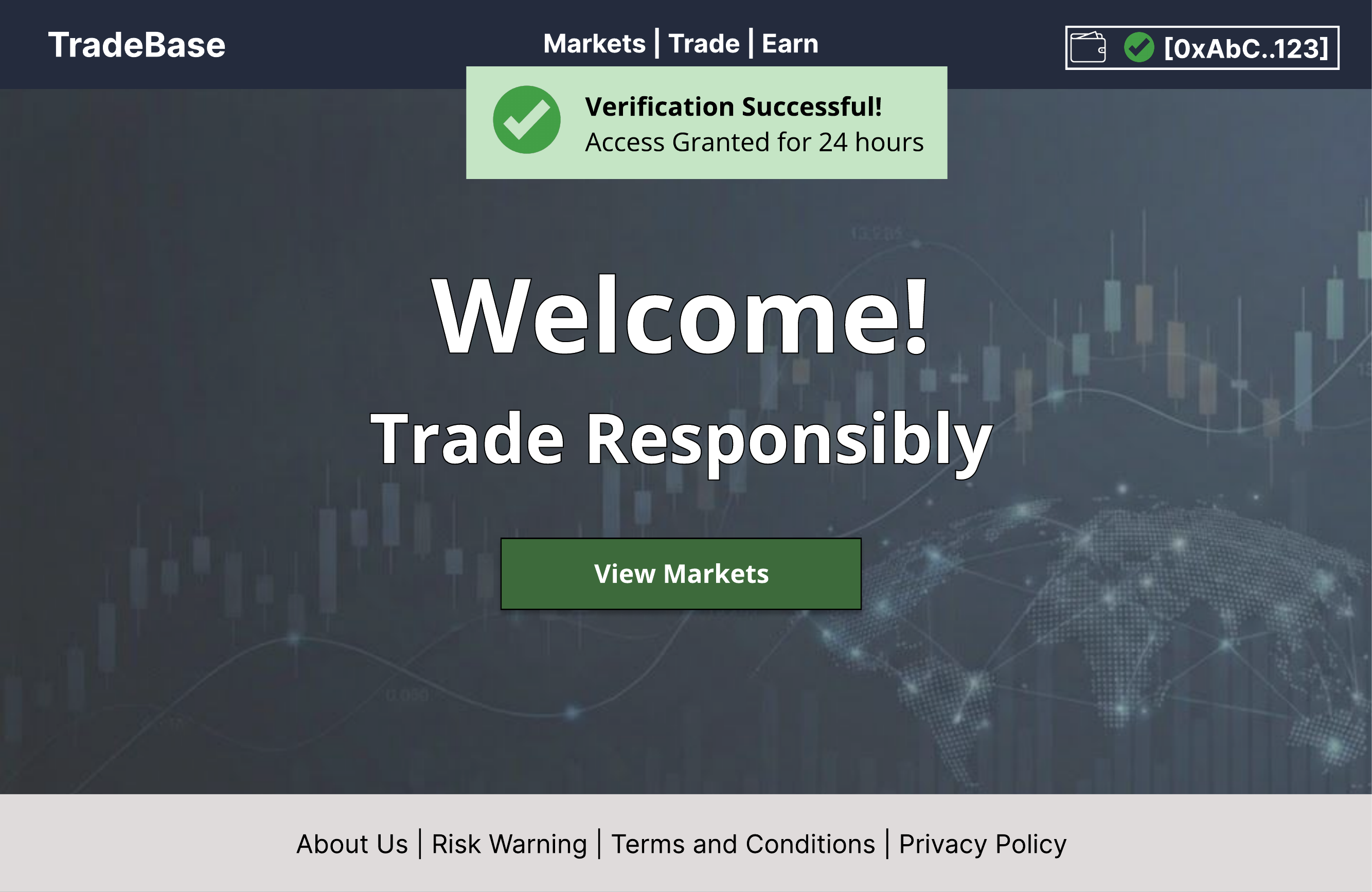}
        \caption{Verification Success \& Entry}
        \label{fig:page4}
    \end{subfigure}
    
    \caption{Integration of the \textbf{ZK-Compliance} protocol within \textbf{TradeBase}, a reference trading application. The sequence shows: (a) the host app enforcing restrictions, (b) the protocol requesting privacy-preserving verification, (c) the local generation of the zk-SNARK, and (d) the verified state granting access.}
    \label{fig:wireframe_grid}
    \Description[A four-step UI sequence showing the ZK-Compliance workflow.]{A grid of four screenshots illustrating the user flow in the TradeBase prototype:
    (a) The application dashboard is locked, displaying an 'Age Restricted' overlay preventing access.
    (b) A ZK-Compliance modal appears, asking the user to prove they are 18+ without revealing their birth date.
    (c) A loading screen shows 'Generating Zero-Knowledge Proof', indicating that cryptographic computation is happening locally on the client.
    (d) The dashboard unlocks with a 'Verified' status, granting the user full access to the trading interface.}
\end{figure*}

\paragraph{\textbf{Phase 2: Verify (Ephemeral Authorization)}}
\textit{Alice} then submits this generated proof to the on-chain system. Upon successful verification, the registry creates an \texttt{AccessRecord} binding her address to the specific service. In a traditional custodial model, \textit{Alice}'s raw data would now reside indefinitely on a corporate server, often without her visibility into its lifecycle. In contrast, our system's \texttt{AccessRecord} functions similarly to a transparent, time-bounded web session cookie stored on the blockchain. 

Crucially, when \textit{Alice} generates the proof, the system enforces a strict expiry timestamp. From \textit{Alice}'s perspective, this transforms an open-ended data handover into a temporary, predictable session. Even if she takes no further action, her permission naturally expires. This provides \textit{Alice} with the peace of mind that her authorization is ephemeral, effectively preventing the accumulation of stale permissions that plague traditional identity systems. For developers, this abstracts cryptographic complexity into a simple logic check, ensuring they operate outside the scope of custodial data regulations like GDPR~\cite{gdpr2016}.

\paragraph{\textbf{Phase 3: Revoke (The ``Kill Switch'')}}
The most significant departure from existing compliance workflows is the introduction of a user-controlled ``Kill Switch.'' If \textit{Alice} decides to stop using a traditional service, her revocation mental model involves navigating complex settings menus or sending a formal deletion request to a support team, with no technical guarantee that her data will actually be purged.

In our framework, revocation is an immediate, cryptographic action. If \textit{Alice} wishes to sever ties with the application, she simply submits a transaction to the registry to delete her \texttt{AccessRecord}. Because the application is required to query this registry for every interaction, \textit{Alice}'s deletion takes effect immediately. The application is cryptographically blocked from verifying her eligibility, instantly terminating the service relationship. This tangible mechanism provides \textit{Alice} with an enforceable exit capability, restoring the balance of power and reinforcing her mental map of absolute control over her digital presence.

\section{ZK-Compliance: Prototype and Evaluation}
To validate the feasibility of the \emph{Grant, Verify, Revoke} lifecycle, we developed \textbf{ZK-Compliance}. To evaluate the system in a realistic environment, we integrated it into \textbf{TradeBase}, a mock fintech reference application designed to simulate standard regulatory friction points (shown in Figure~\ref{fig:wireframe_grid}). This full-stack prototype targets \textbf{Age Verification}, a use case that represents a major barrier in the current Web3 ecosystem. By embedding the privacy protocol directly into TradeBase, we can demonstrate exactly how the user's mental model shifts when they are given cryptographic agency at a compliance gate.

\subsection{Implementation: Architecture and Usability}
To avoid reliance on trusted backends and to reduce user friction, we utilize browser local storage as a non-custodial Identity Vault. We used Circom~\cite{belles2022circom} and SnarkJS~\cite{iden3snarkjs} to handle witness generation and proving entirely within the client's JavaScript environment. From an interaction perspective, this means the user never has to leave the application context or be redirected to a third-party portal to authenticate. This standardized tooling ensures the system is portable across any modern web browser without requiring users to install specialized wallet extensions or understand complex key management. The cryptographic heavy lifting is abstracted entirely behind a familiar loading screen. A major usability challenge in privacy systems is protecting simple data (like a birth year) without burdening the user with extra steps~\cite{das2024design,saka2025sok,saka2024evaluating,tazi2024evaluating,kishnani2024towards,das2022privacy,das2024user}. To protect low-entropy attributes from brute-force guessing, the system mathematically binds a local, high-entropy secret to the data before generating the proof. Because this secret is generated and stored automatically within the local vault, the user is never asked to manage, memorize, or input it. This prevents attackers from reverse-engineering the attribute without physically compromising the user's device~\cite{toutsop2021exploring,kishnani2025security}, ensuring strong privacy guarantees are delivered through a frictionless, single-click interface.

\subsection{Evaluation: Interaction Latency and Economic Feasibility}
For privacy tools to achieve widespread adoption, latency is a crucial factor; high friction drives users toward insecure alternatives. Prior work establishes that Web3 usability hinges on the user's ability to exercise effective control despite technical complexity~\cite{guan2025using,gupta2024really}. Consequently, we prioritized~\textit{interaction latency} (the delay between a user's intent to prove eligibility and the cryptographic execution) as a primary design constraint. We evaluated the system on standard consumer-grade hardware to ensure accessibility. The local circuit achieves client-side proof generation in~\textit{under 200~ms}. This result is significant as it shows that cryptographic privacy does not require a trade-off with responsiveness. The delay is imperceptible, allowing these privacy checks to be integrated seamlessly into standard login flows without degrading the user experience. 

Beyond performance, we analyzed on-chain verification costs to assess economic accessibility, as high fees can act as an exclusionary barrier. On the Ethereum Mainnet, this resulted in a prohibitive cost of approximately $\$15$ at standard gas prices (using a representative gas price of 20~gwei and ETH $\approx\$3{,}000$ under typical network conditions). However, based on current fee structures for Layer 2 scaling solutions (e.g., Optimism or Arbitrum), we estimate that the cost for the same verification workload would drop to under $\$0.50$. This projection suggests that while the interaction model is sound, deploying on scalable infrastructure is a necessary prerequisite for democratizing access to privacy-preserving compliance tools.

\section{Discussion: Enforcing Digital Rights}
We analyze the proposed framework not merely as a security protocol, but as a mechanism for enforcing digital rights. By embedding privacy protections directly into the interaction layer, the system addresses the structural power imbalances inherent in current compliance workflows.

\textbf{Minimization and Surveillance Defense.} Standard verification workflows often create a ``honey pot'' risk by permanently recording sensitive attributes in centralized databases or immutable ledgers. Our system counters this by enforcing strict \textit{data minimization}. By recording only a boolean verification signal and an opaque cryptographic proof, we decouple access rights from identity. This effectively instantiates a technical ``right to be forgotten'' even within an immutable ledger environment. To further protect users from brute-force attacks on low-entropy attributes (e.g., birth years), we mandate the inclusion of a high-entropy \texttt{randomSalt} (128 bits) within the user's local vault. This salt is mathematically bound to the attribute before proof generation, ensuring that private data cannot be reverse-engineered via enumeration. This architecture fundamentally shifts the attack surface: in a centralized model, a single breach compromises all users, whereas in our client-side model, an attacker would need to compromise individual user devices one by one to access raw attributes. This distributes risk to the edges, making mass surveillance and bulk data exfiltration economically impractical.

\textbf{Sovereign Revocation.} A defining feature of our framework is the implementation of a user-controlled ``Kill Switch''. In centralized systems, revocation is typically an administrative request: the user asks a server to delete their data, but must trust the provider to honor that request and purge cached credentials. In our framework, revocation is a cryptographic state change. When a user invokes the revocation function, their specific \texttt{AccessRecord} is deleted from the on-chain registry. Because the application logic requires this record to proceed, the revocation takes immediate effect. The application cannot verify the user anymore, regardless of its internal policy. This enforces a non-cooperative revocation model where the user does not need the service's permission to terminate the relationship. The deletion of the on-chain state cryptographically invalidates the verification path, ensuring that the user's decision to exit is final, immediately effective, and globally consistent.

\section{Future Work and Limitations}
The current prototype has some limitations to address before becoming a production-ready system. Its reliance on self-attestation restricts its utility in regulated sectors requiring facts from trusted authorities. Additionally, the system presents a security vulnerability because verification proofs on the public network are not strictly bound to specific user identities, creating risks of interception and reused consent. The prototype also depends on a developer-controlled initialization process, which contradicts the principles of decentralized consensus. Finally, the cognitive load of navigating the Grant, Verify, Revoke compliance model and the users' ability to confidently exercise revocation remain empirically untested.

To resolve these constraints, future work will focus on integrating high-assurance credentials. This will allow users to seamlessly import authority-issued facts into local privacy vaults while cryptographically stripping all Personally Identifiable Information before ledger submission. We will also harden the protocol to bind eligibility proofs tightly to user identities, ensuring actions are unforgeable and consent cannot be hijacked. Furthermore, we plan to transition to a transparent, community-driven initialization process governed by the users. Finally, we will conduct comprehensive empirical studies to validate the human-centered security design and measure how effectively users navigate their privacy boundaries.

\section{Conclusion}
The tension between regulatory compliance and user privacy in Web3 environments requires a fundamental shift in how systems handle digital identity. In this paper, we presented a \textit{Selective Disclosure Framework} that shifts identity verification from a permanent data handover into a dynamic, user-controlled lifecycle. By utilizing client-side Zero-Knowledge Proofs through our \textit{ZK-Compliance} prototype, we demonstrated that users can satisfy strict eligibility requirements without surrendering their underlying data. The introduction of the \emph{Grant, Verify, Revoke} model, specifically the cryptographic ``Kill Switch,'' successfully restores user agency and enforces the principle of data minimization at the architectural level. Hence, this framework shows that radical transparency and regulatory compliance do not have to come at the cost of self-sovereignty. As blockchain applications continue to integrate with regulated sectors, empowering users with cryptographic control over their own disclosures will be essential for building a truly secure and equitable decentralized web.

\begin{acks}
    We acknowledge the Data Agency and Security (DAS) Lab at George Mason University, where this study was conducted. We also thank Dr. Xiaokuan Zhang for initial feedback on this work. This research was supported in part by an unrestricted gift from Google. The opinions expressed in this work are solely those of the authors.
\end{acks}

\bibliographystyle{ACM-Reference-Format}
\bibliography{main}

\end{document}